%% file: main.tex
\documentclass[%
 reprint,
 amsmath,amssymb,
 aps,
prx,
longbibliography,
]{revtex4-1}

\usepackage{graphicx}
\usepackage{dcolumn}
\usepackage{bm}
\usepackage{mathtools}
\usepackage{graphicx} 
\usepackage{comment}

\DeclareUnicodeCharacter{2212}{-}
\begin{document}

\preprint{APS/123-QED}

\title{Dynamic Surfactants Drive Anisotropic Colloidal Assembly 
}

\author{Yaxin Xu}
\author{Prabhat Jandhyala}
\author{Sho C. Takatori} %
 \email{corresponding author: stakatori@ucsb.edu}
\affiliation{%
 Department of Chemical Engineering, University of California, Santa Barbara, Santa Barbara, California, USA 93117.
}%

\date{\today}

\begin{abstract}
\input{abstract}
\end{abstract}

\maketitle

\section{Introduction}
\input{introduction}

\section{Methods}
\input{methods}

\section{Results}
\input{results}

\begin{acknowledgments} 
This material is based upon work supported by the Air Force Office of Scientific Research under award number FA9550-21-1-0287.
Y.X. acknowledges support from the Dow Discovery Fellowship at UC Santa Barbara. 
S.C.T. is supported by the Packard Fellowship in Science and Engineering.
Acknowledgment is made to the Donors of the American Chemical Society Petroleum Research Fund for partial support of this research.
Use was made of computational facilities purchased with funds from the National Science Foundation (OAC-1925717) and administered by the Center for Scientific Computing (MRSEC; NSF DMR 1720256) at UC Santa Barbara.
\end{acknowledgments}

\bibliographystyle{apsrev4-1}
\bibliography{bibliography.bib}
\end{document}

%% file: abstract.tex
Colloidal building blocks with re-configurable shapes and dynamic interactions can exhibit unusual self-assembly behaviors and pathways.
In this work, we consider the phase behavior of colloids coated with surface-mobile polymer brushes that behave as ``dynamic surfactants.'' 
Unlike traditional polymer-grafted colloids, we show that colloids coated with dynamic surfactants can acquire anisotropic macroscopic assemblies, even for spherical colloids with isotropic attractive interactions.
We use Brownian Dynamics simulations and dynamic density functional theory (DDFT) to demonstrate that time-dependent reorganization of the dynamic surfactants leads to phase diagrams with anisotropic assemblies.
We observed that the microscopic polymer distributions impose unique geometric constraints between colloids that control their packing into lamellar, string, and vesicle phases.
Our work may help discover versatile building blocks and provide extensive design freedom for assembly out of thermodynamic equilibrium.

%% file: introduction.tex
In colloidal self assembly, the shape of particle building blocks may be engineered to achieve specific interparticle interactions and pack into well-defined, complex geometries.
Understanding the interplay between particle shape and structures is essential for designing advanced colloidal materials with tailored properties.
In general, spherical colloids with isotropic attractive interactions associate into a limited set of structures with space-efficient unit cells, such as HCP and FCC, once the interaction strength exceeds the energy of thermal fluctuations.\cite{Glotzer2004}
For anisotropic colloids that interact directionally along different axes, the phase diagram contains more open and diverse morphologies that are inaccessible by spherical colloids with isotropic interactions.\cite{Sacanna2011,Manoharan2015}
For example, patchy DNA-coated colloids may assemble into strings and sheets by limiting the patch coordination number or valency.\cite{Sacanna2013, Anders2014, Chen2019, Wang2012, Zhang2021,Angioletti-Uberti2014, Romano2011} 
Elongated, rod-like colloids promote multi-body alignment and stacking, favoring the formation of aligned liquid crystalline phases and colloidal membranes.\cite{Onsager1949, Barry2010}
As another example, recent works have shown that shape-complementary colloidal pairs with cavities and internal voids may be exploited for lock-and-key self assembly mechanisms.\cite{Sacanna2013,Cinacchi2010} 

While anisotropic colloids formed through irreversible chemical and physical modifications have been well studied, colloids that can reversibly reconfigure their isotropic-to-anisotropic interactions offer new avenues for self assembly.
Uniformly grafted nanoparticles can form anisotropic assemblies such as sheets and strings, although the physical mechanisms for this behavior have not been identified.\cite{Akcora2009}
Colloids with dynamic shapes could exhibit adaptive behaviors such as structure reconfiguration or initiation of specific assembly pathways on demand, thereby avoiding kinetic traps that lead to uncontrolled aggregation and structural defects.
To achieve such dynamic interactions, we have demonstrated in a previous paper the capability to engineer colloidal particles with surface-mobile, sterically-hindering polymer brushes.\cite{Xu2023}
Surface-mobile polymers enable a dynamic, contact time-dependent pair interaction which may be regulated by nonequilibrium forces.
Other groups have shown that colloids coated by surface-mobile binders such as DNA-linkers are able to reversibly form adhesive patches, enabling multi-stage self assembly approaches.
\cite{Leunissen2009,VanDerMeulen2013,Angioletti-Uberti2014, Mitra2023} 
By harnessing the inherent responsiveness of these dynamic particles, we aim to engineer colloids with surface-tunable properties that are responsive to changes in their environment. 

\begin{figure}
\centering
\includegraphics[width=0.47\textwidth]{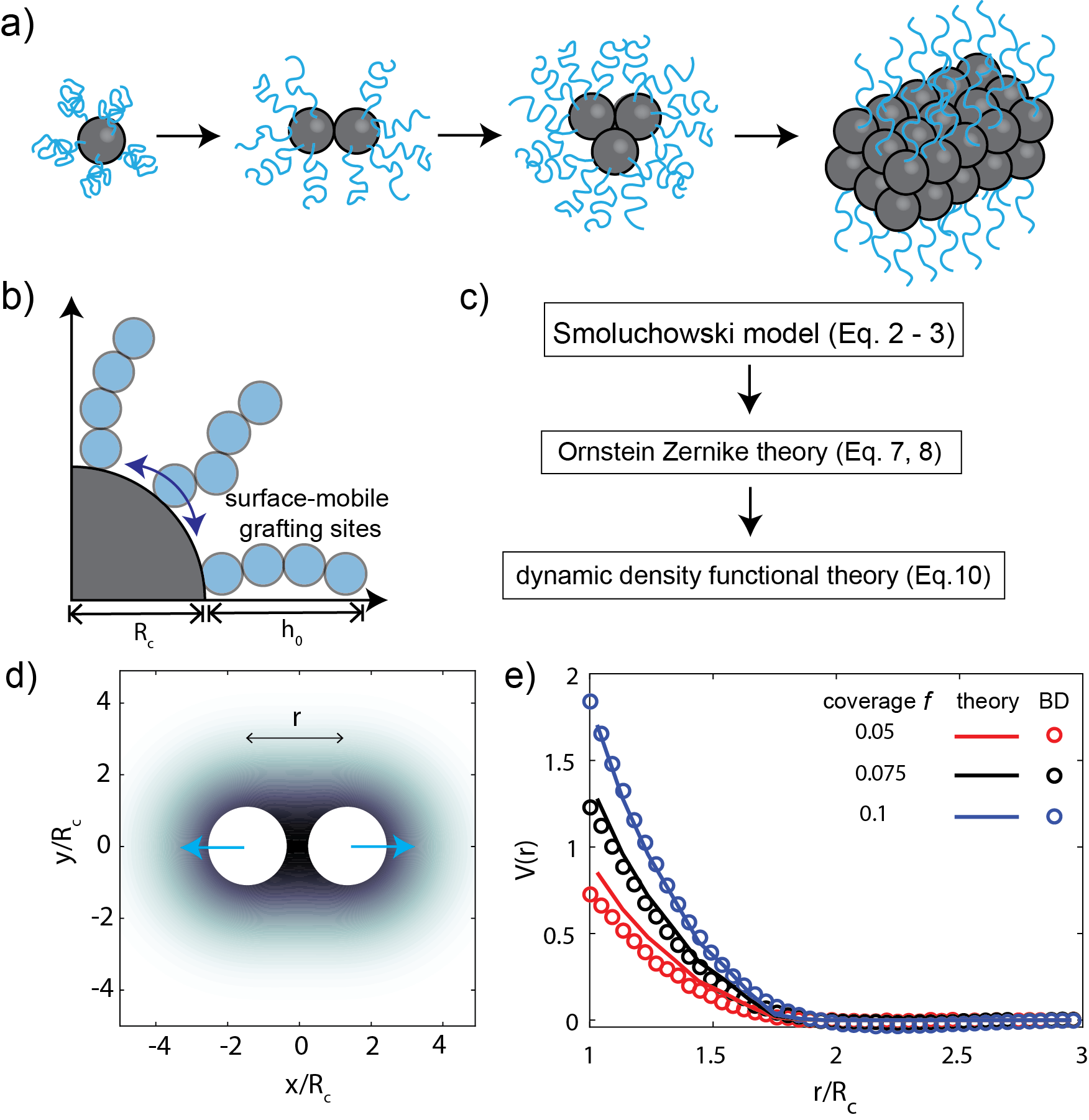}
\caption{Multi-body interactions between attractive colloids coated with ``dynamic surfactants'' induce anisotropic assemblies.
A) Cartoon showing progression of self assembly from single colloids (gray) coated by surface-mobile polymers (blue).
Polymers exclude out of the contact interfaces to minimize steric interactions, resulting in anisotropic growth of superstructures.
B) Schematic of end-grafted polymers whose grafting sites are free to diffuse laterally along the surfaces, with average brush height $h_{\text{0}}$ coating the colloid of radius $R_{\text{c}}$. 
C) Flow diagram of our theory that couples the polymer conformations to colloidal phase behavior.
D) Cross-sectional plot showing the solution to Eqs.~\ref{Smoluchowski}-\ref{Smoluchowski_Flux} for polymer distribution $\rho$ for a pair of colloids separated at distance $r$.
E) Potential of mean force from Eq.~\ref{PMF} as a function of $r$ with Brownian dynamics (BD) simulation results (markers) and theoretical predictions (line) for three different surface coverages.
BD data is obtained through a Boltzmann inversion of a semi-dilute suspension. 
}
\label{Fig1}
\end{figure}

In this paper, we use Brownian Dynamics (BD) simulations and liquid state theory to study the self assembly of colloids coated with surface-mobile polymers that act as ``dynamic surfactants'' (Fig.~\ref{Fig1}a-b).
Multibody effects drive the reorganization of local surface polymers such that intrinsically isotropic colloids may reversibly adopt anisotropic patches upon assembly into higher order structures.
By controlling density, attraction strength, and polymer surface coverage, we observed a rich self assembly phase space that is inaccessible by colloids with purely isotropic interactions, including vesicles, bilayers, and strings.
We find that surface-mobile polymers are analogous to amphiphilic surfactants, where a critical packing parameter relating the head and tail shapes predicts the formation of micelles and bilayers.\cite{Israelachvili2010}

%% file: methods.tex
\begin{figure*}
\centering 
\includegraphics[width=0.85\textwidth]{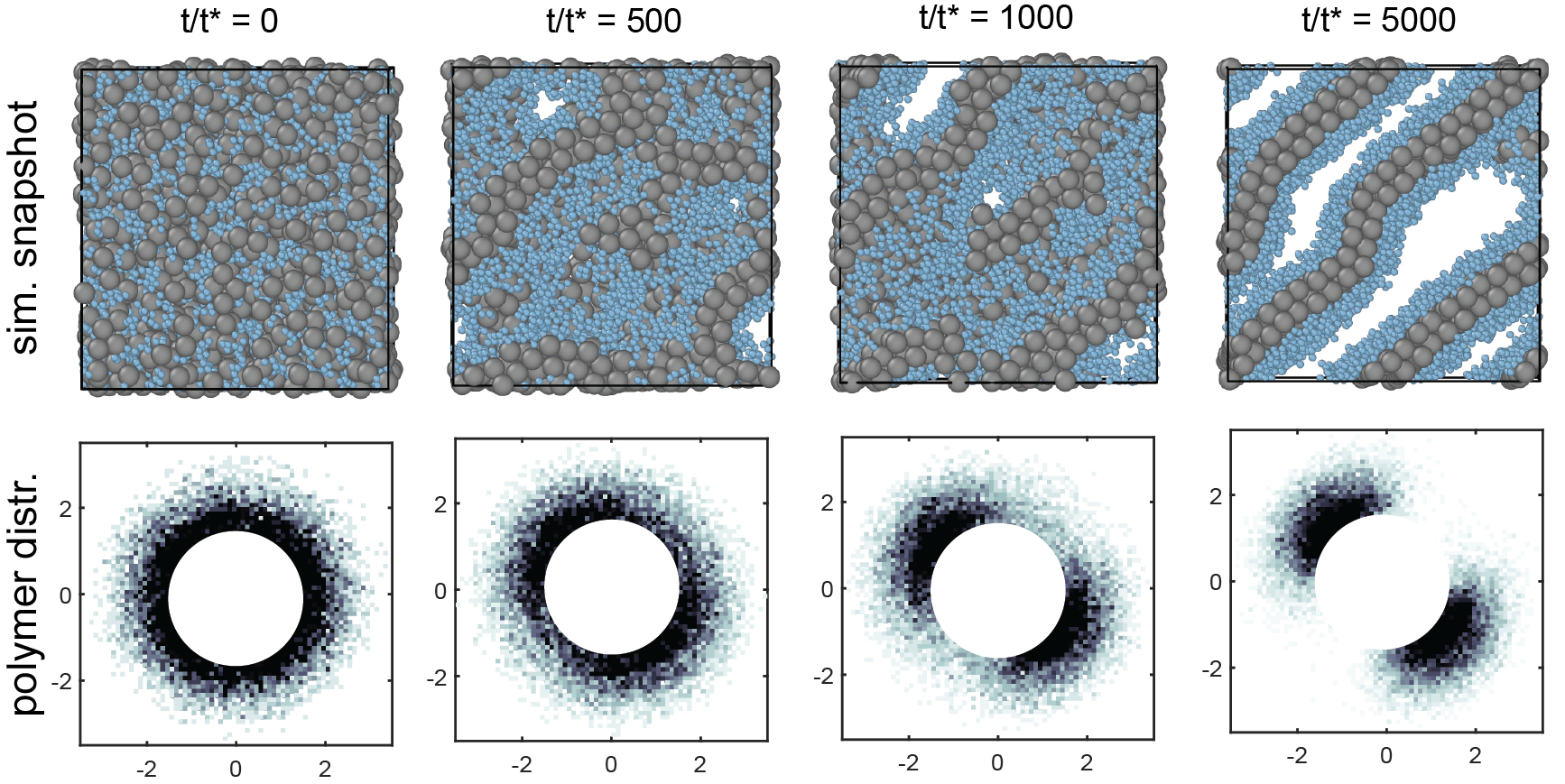}
\caption{
Macroscopic colloidal self assembly is coupled to surface rearrangements of the dynamic surfactants.
The BD snapshots of polymer-coated colloids (top row) and corresponding averaged polymer distributions viewed along one direction (bottom row) are shown as a function of increasing simulation time (left to right).
At short times $t/t^* \leq 500$, colloids associate to form an intermediate string phase and do not exhibit global structure.
The corresponding average polymer distribution remains uniformly distributed in a corona around the colloids.
At longer times, the colloidal assemblies rearrange to form a lamellar phase consisting of bilayer sheets that stack together.
Polymers visibly segregate to above or below the colloidal sheet to minimize the steric overlap at colloid-colloid contact sites.
Colloidal packing fraction is $\phi = 0.25$, surface coverage is $f = 0.05$, chain length is $m = 2$, and reduced temperature is $k_{\text{B}}TR_{\text{c}}/a = 0.24$.
The characteristic Brownian timescale of the colloid is $t^{*} = 4R_{\text{c}}^2/D_{\text{c}}$.
}
\label{Fig2}
\end{figure*}

\subsection{Simulation protocol}
As shown in Fig.~\ref{Fig1}b, we perform Brownian Dynamics (BD) simulations of colloids of radius $R_{\text{c}} = \sigma$ coated by surface-mobile polymers and suspended in an implicit Newtonian solvent of viscosity $\eta$.
We briefly summarize the protocol for tethering polymer chains onto colloidal surfaces, as documented previously.\cite{Xu2023}
Surface-mobile polymers are modeled using a semi-flexible Kremer-Grest bead-spring model, where the bead diameter is $d_{\text{p}} = 0.7\sigma$.
Each polymer chain comprises $n$ polymer beads and each colloid comprises $m$ monomers.\cite{Kremer1990}
In Fig.~\ref{Fig1}c, one end bead of the polymer is constrained on the colloidal surface via a stiff harmonic bond such that it may freely diffuse tangentially but cannot detach from the surface.
All non-bonded particle pairs interact through a Weeks-Chandler-Anderson (WCA) hard sphere-like potential.
To study self assembly, we introduce a short-range attraction between the colloidal cores via a Yukawa potential with magnitude $-a$, screening length $\kappa = 1$, and truncation at $r_{\text{cut}} = 3R_{\text{c}}$.

All particles in the simulation follow the overdamped Langevin equations of motion:
\begin{equation}
\zeta \frac{\Delta \textbf{x}_{i}}{\Delta t} = \quad \underbrace{\textbf{F}_{i}^{\text{P}}}_{\mathclap{\text{interaction}}} + \quad \underbrace{\textbf{F}_{i}^{\text{R}}}_{\mathclap{\text{thermal}}}
\label{Eq3}
\end{equation}
with contributions from interparticle interactions and thermal forces that satisfy fluctuation dissipation.
In Eq.~\ref{Eq3}, $\zeta$ is the drag coefficient following the Stokes-Einstein-Sutherland relation.
For all simulations, we choose a time step $\Delta t = (2 \times 10^{-4})R_{\text{c}}^2/D_{\text{c}}$ and sample the simulation every $10^5$ time steps.
The characteristic timescale $t^{*} = 4R_{\text{c}}^2/D_{\text{c}}$ describes the Brownian timescale of the colloid to diffuse the distance of its diameter.
To obtain representative statistics, we simulate $N_{\text{c}} = 1000-1200$ colloidal particles for $8 \times 10^{8}$ time steps.
Due to the coarse-graining of polymer segments, our simulation results capture relaxation beyond the Brownian timescale of the polymer bead $t\sim d_{\text{p}}^2/D_{\text{p}}$ and do not account for fluctuations at smaller length and timescales.
Simulations are performed in a 3-dimensional box of volume $L^3$, where $L$ is varied between $20\sigma$ and $60\sigma$ to produce the desired packing fraction, using HOOMD-Blue, a GPU-accelerated Python package.\cite{Anderson2020} 

\subsection{Theoretical model}
To complement our BD simulations, we develop a theoretical framework to understand how interactions facilitated by surface-mobile polymers influence macroscopic colloidal phase behavior.
As outlined in Fig.~\ref{Fig1}c, we use a field-based Smoluchowski theory to model polymers and extract a potential of mean force experienced by pairs of colloids. 
We then compute the correlation functions in the bulk fluid using a standard liquid state approach.\cite{Hansen2013}
Finally, we apply dynamic density functional theory (DDFT) to predict the phase behavior and stability of the system.

The analytical theory to capture the effective colloidal interaction mediated by surface-bound polymers has been documented \cite{Xu2023}. 
Assuming that the surface polymers relax quickly relative to the timescale at which colloids self-assemble, the probability density $\rho(\textbf{h},t)$ of finding a polymer bead at position $\textbf{h}$ satisfies the steady state Smoluchowski equation:
\begin{equation}\label{Smoluchowski}
0 = -\nabla \cdot \textbf{j}(\textbf{h})
\end{equation}
where the nondimensional polymer flux contains thermal, entropic elasticity, and excluded volume contributions:
\begin{equation}\label{Smoluchowski_Flux}
\textbf{j}(\textbf{h}) = - \nabla \rho(\textbf{h}) - \rho(\textbf{h}) \nabla  \frac{U_{\text{el}}(\textbf{h})}{k_{\text{B}}T} - \rho(\textbf{h}) \nabla \frac{U_{\text{excl}}(\textbf{h})}{k_{\text{B}}T}.
\end{equation}
In Eq.~\ref{Smoluchowski} - \ref{Smoluchowski_Flux}, we have nondimensionalized time by $t^*$ and all distances by $d_{\text{p}}$.

The excluded volume contribution is given by a nonlinear term:
\begin{equation}\label{U_excl}
    \frac{U_{\text{excl}}(\textbf{h})}{k_{\text{B}}T} = B_{2} \int \rho(\textbf{h}) \rho (\textbf{h} - \textbf{h}') d\textbf{h}'
\end{equation}
where $B_{2}$ is the excluded volume parameter.
Using well-established polymer brush theory, we assume that the entropic elasticity of the polymer brush around a single colloid can be described by a harmonic potential of the form $U_{\text{el}} \sim \frac{r^2-(R_{\text{c}}+\frac{d_{\text{p}}}{2})^2}{n^2 d_{\text{p}}^2}$ which penalizes polymer beads from being strongly stretched \cite{DeGennes1987,Milner1991}.
We approximate the external potential for polymer beads around two colloids by inverting the sum of two Boltzmann distributions about each colloid:
\begin{equation}\label{U_el}
    \frac{U_{\text{el}}(\textbf{h})}{k_{\text{B}}T}= -\text{log} \left[ e^{-\frac{r_1^2-(R_{\text{c}}+\frac{d_{\text{p}}}{2})^2}{n^2 d_{\text{p}}^2}} + e^{-\frac{r_2^2-(R_{\text{c}}+\frac{d_{\text{p}}}{2})^2}{n^2 d_{\text{p}}^2}} \right].
\end{equation}
where $r_1(\textbf{h}),r_2(\textbf{h})$ are the distances from each of the colloid centers.
Furthermore, the system obeys mass conservation $\int \rho dr^3 = nm$ and satisfies no flux boundary conditions on the colloidal contact interfaces $S_{1},S_{2}$ as well as zero density at $r \rightarrow \infty$.
Equations \ref{Smoluchowski}-\ref{Smoluchowski_Flux} are numerically evaluated using the finite element software package FreeFEM++ for an arbitrarily-large 3-dimensional volume which includes both colloidal particles and the two polymer brush domains \cite{Hecht2012}.
In Fig.~\ref{Fig1}d, we show the solution $\rho$ for a given colloidal separation of $r/R_{\text{c}} = 1.5$.
The higher density of polymers at the interface results in a repulsive force that pushes the colloids apart.

From the polymer distribution $\rho$, we may compute a potential of mean force $V(r)$ for two colloids, separated at $r$, to overcome the osmotic force exerted by polymers along the colloidal pair's line of centers \cite{Loverso2012}.
Since we only account for hard-sphere interactions, this potential reduces to a simple form:
\begin{equation}
    \label{PMF}
    \frac{V(r)}{k_{\text{B}}T} = \int_{\infty}^{r} \oint_{\partial S} \rho (r')d\textbf{S} dr' 
\end{equation}
where \textbf{S} is the contact surface between the polymers and the colloid.

In Fig.~\ref{Fig1}e, we present $V(r)$ as a function of the separation $r$ for three different surface coverages $f$.
The BD simulations are obtained through Boltzmann inversion of the pair distribution function in a dilute system without attraction.
This calculation is analogous to integrating the force experienced by a single colloid at equilibrium when separated from another colloid at distance $r$.
In Fig.~\ref{Fig1}e, we observe that the theoretical prediction  agrees well with BD results for all surface coverages, but slightly overestimates the near-field repulsion at $1 \leq r/R_{\text{c}} \leq 1.5$.
This overestimate is attributed to the inaccuracy of the brush potential in describing the polymers at such dilute densities, where it is expected that the surface-tethered polymers behave in a mushroom regime as opposed to brush regime.
Due to the mass conservation constraint, the potential of mean force is proportional to the number of polymer chains $m$ and consequentially the surface coverage $f$.

\subsection{Dynamic Density Functional Theory}
In order to predict the phase behavior of our colloids with dynamic surfactants, we turn towards spinodal decomposition theory.\cite{Caccamo1993,Mahynski2015}
First, we use liquid state theory to extract equilibrium correlation functions for our inhomogenous system.
The equilibrium structural correlations are well described by the Ornstein-Zernike equation,
\begin{equation}\label{OZ}
    h(r) = c_{\text{0}}(r) + \psi_{0} \int c_{\text{0}}(r-r') h(r') d r'
\end{equation}
where $\psi_0$ is the average number density, $h(r) = g(r)-1$ is the total correlation function, and $c_{\text{0}}(r)$ is the direct correlation function between polymer-coated colloids without attraction.
To solve Eq.~\ref{OZ} for the two unknown functions $h$ and $c_{\text{0}}$, we require a closure relation.
For simplicity, we choose the hypernetted chain (HNC) closure:
\begin{equation}\label{HNC}
h(r) = e^{-\beta V(r) + h(r) - c_{\text{0}}(r)} -1.
\end{equation}

To evaluate Eq.~\ref{OZ}, we use fast Fourier transforms (FFT) and choose $1024$ grid points with a domain size of $L = 20 \pi$.
We use a standard Picard iteration scheme to evaluate the equations, \cite{Rogers1980,Rogers1984} and converge solutions until we reach a tolerance of $10^{-14}$.
Note that the potential of mean force $V(r)$ does not contain the attractive component and is purely the polymer-mediated repulsion.

We next invoke the random phase approximation and assume that the direct correlation function $c(r)$ of the corresponding equilibrium fluid with attraction is well described by:
\begin{equation}
    c(r) = c_{\text{0}} (r;\psi_0) + \beta v_{\text{Yukawa}}(r).
\end{equation}

With the full $c(r)$, we are now in a position to write down the dynamic density functional theory (DDFT) equation for colloidal particles: \cite{Archer2004,Marconi1999}
\begin{equation}\label{DDFT}
    \frac{\partial \psi (\textbf{r},t)}{\partial t} = \nabla \left[ \psi(\textbf{r},t) \nabla \frac{\delta F[\psi(\textbf{r},t)]}{\delta \psi (\textbf{r},t)} \right]
\end{equation}
where the Helmholtz free energy functional $F[\psi(\textbf{r},t)]$ is given by:
\begin{equation}\label{Free_Energy}
    F[\psi({\textbf{r},t})] = F_{\text{id}}[\psi(\textbf{r},t)] + F_{\text{ex}}[\psi(\textbf{r},t)].
\end{equation}
In Eq.~\ref{Free_Energy} the first term represents the ideal gas entropic contribution $F_{\text{id}} = k_{\text{B}}T \int d\textbf{r} \psi(\textbf{r},t) [\text{ln}(\psi(\textbf{r},t) \Lambda^3) -1 ]$ and the second term is the excess free energy that models interactions with other colloids in the fluid.
The DDFT approach relies on the adiabatic approximation which assumes that the nonequilibrium system evolves based on equilibrium density correlations.\cite{Hansen2013}

%% file: results.tex
\subsection{``Dynamic surfactant'' rearrangements}
\begin{table}
\centering
\includegraphics[width=0.47\textwidth]{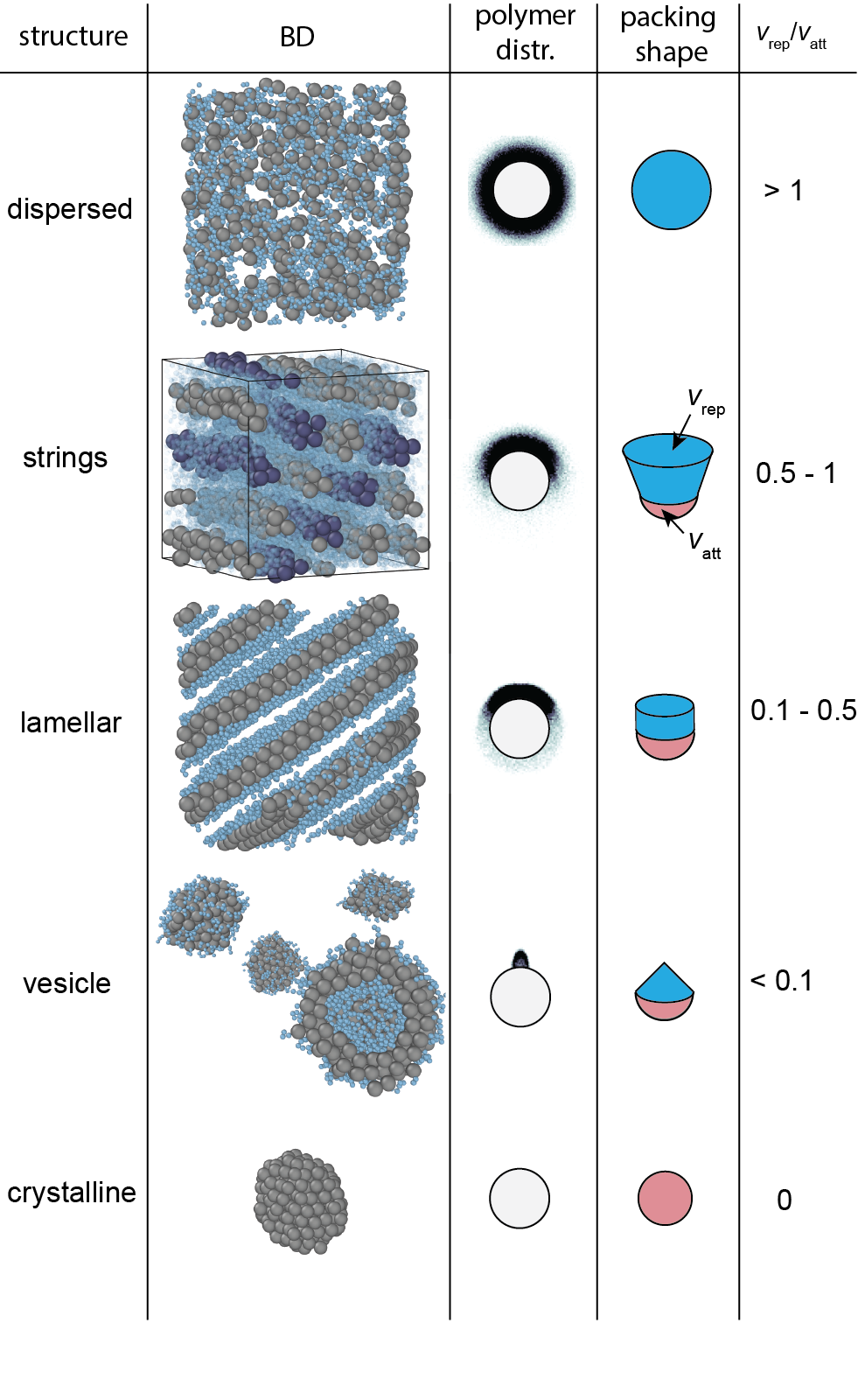}
\caption{Colloids with dynamic surfactants assemble into anisotropic structures.
Second column is the snapshot from BD simulations, third column is the local average polymer distribution from simulations, and fourth column is a cartoon of the packing shape of the colloidal building block, with the attractive core in red and the sterically-hindering polymer layer in blue. 
Last column indicates the packing parameter which is a ratio of the excluded volume by the polymer beads $v_{\text{rep}} = nm \pi d_{\text{p}}^3/6$ and the volume of the attractive core $v_{\text{att}} = \pi 4R_{\text{c}}^3/3$.
The reduced temperature is $k_{\text{B}}T R_{\text{c}}/a = 0.2$ for all structures.
}
\label{Table1}
\end{table}

We first consider the temporal evolution of the colloidal structure and local polymer configurations.
Simulations of colloids with isotropic attractive interactions and repulsive surface-mobile polymers were performed at packing fraction $\phi = \frac{N_{\text{c}} \pi 4 R_{\text{c}}^3}{3L^3} = 0.25$, fractional surface coverage $f = A_{\text{p}}/A_{\text{c}}=0.05$ where $A_{\text{c}} = \pi R_{\text{c}}^2$ is the colloid surface area, $A_{\text{p}} = m\pi R_{\text{g}}^2$ is the estimated colloidal surface area covered by polymers, and $R_{\text{g}} = nd_{\text{p}}^2/6$ is the radius of gyration of the linear chain.\cite{Doi1988}
In Fig.~\ref{Fig2}, we present both the simulation snapshot and heat maps of the corresponding polymer distribution around individual colloids, averaged over the entire suspension and viewed along one direction. 
The suspension begins from a homogeneous, dispersed state at $t/t^{*} = 0$ where polymer chains appear spatially uniform and evenly distributed around the colloids, resulting in isotropic interactions.
At intermediate times, the colloids spontaneously self assemble into string-like aggregates that are 2-3 particles wide to minimize their free energy while maximizing polymer entropic degrees of freedom.

At late stages of the assembly process, $t/t^{*} \geq 1000$, we observe a transition from string-like structures to a lamellar phase, consisting of bilayer colloidal sheets stacked together.
Correspondingly, surface polymers segregate into two distinct poles and expose the colloid-colloid contact areas, consistent with the multi-step assembly pathway hypothesized in Fig.~\ref{Fig1}a.
The resulting anisotropy is an entropic effect unique to our dynamic surfactants in contrast to traditional polymer-coated particles where the grafting sites are fixed in place.
The bilayer structure resembles prior work on patchy particles, despite the fact that the surface-mobile polymers assume isotropic distributions on an isolated colloid.\cite{Miller2009,Preisler2014}
The polymer densities at the poles experience up to a $4x$ increase, which acts to swell the polymer brushes and limit colloidal assembly along the perpendicular direction to the bilayer.
Additionally, the presence of the polymer brush exerts a lateral osmotic pressure which stabilizes the bilayer against collapse, similar to how incorporation of polymer-grafted lipids increases the bending rigidity and elastic modulus of lipid bilayers.\cite{Hristova1994}
Note that while in simulations the polymers exclude to either above or below the bilayer, Fig.~\ref{Fig2} shows polymer density at both hemispheres due to averaging.

\subsection{Critical polymer packing determines structure}
We have shown in the previous section that macroscopic colloidal assembly is dynamically coupled to microscopic polymer conformations.
To quantitatively understand this relationship, we investigate the steady state structures achievable through tuning the polymer surface coverage.
Here, the steric hindrance imposed by the polymers may be quantified by a critical packing parameter, which is a ratio of the polymer excluded volume $v_{\text{rep}} = nm \pi d_{\text{p}}^3/6$ and the volume of the attractive core $v_{\text{att}} = 4 \pi R_{\text{c}}^3/3$.
By varying $v_{\text{rep}}/v_{\text{att}}$, a quantity which is intrinsic to the polymer properties, we hypothesize that the morphology of the steady state structure may be precisely tuned.

We present resulting superstructures as a function of the critical packing parameter in an organized chart in Table~\ref{Table1}.
When no polymers are present, i.e. $v_{\text{rep}}/v_{\text{att}} = 0$, the colloids form equilibrium HCP crystals at lower packing fractions or kinetically arrest into gel-like aggregates at higher concentrations.\cite{Segre2001, Zia2014}
When surface-mobile polymers are present, different structures arise due to changes in the polymer distribution.
For $v_{\text{rep}}/v_{\text{att}} < 0.1$, we observe formation of vesicle-like structures, where two leaflets of colloids form a spherical shell and polymers are excluded to either the inner or outer surfaces.
We hypothesize that the vesicle state is possible at low $v_{\text{rep}}/v_{\text{att}}$ because the small amount of polymers present on the surface enforces an anisotropic packing shape and prevents collapse into a tighter structure.
Provided that the reduced temperature is sufficiently high, the vesicle is formed from an initial, fluctuating bilayer that spontaneously acquires sufficient membrane curvature.
Note that sufficiently large simulation boxes are required to observe vesicles as opposed to the system-spanning, lamellar structures.\cite{Cooke2005}

At higher $0.1<v_{\text{rep}}/v_{\text{att}}<0.5$, the steric polymer interactions increase the bending energy of the bilayer, such that the lamellar phase becomes preferred over vesicle-type structures.\cite{Hristova1994}
By controlling the strength of the cohesive attraction, the lamellar phase either undergoes spatial fluctuations at higher reduced temperatures or remains rigid as a crystalline layer at lower reduced temperatures.
The orientational ordering and bilayer fluctuations in these simulations qualitatively resemble that of lipid membranes \cite{Yuan2010,Noguchi2011,Shiba2011,Cooke2005} but do not require angle-dependent simulation potentials or hydrophobic and hydrophilic interactions.
We note that to form these structures, the polymers must be able to cylindrically pack onto one hemisphere of the colloidal core, as indicated by the polymer distribution.
At packing parameters $v_{\text{rep}}/v_{\text{att}}$ approaching unity, the bilayers undergo a 2D to 1D phase transition to form coiled, string-like phases that span the simulation box.
We hypothesize that this transition occurs because the polymer entropic contributions exceeds that of the cohesive enthaplic forces which stabilize the bilayer.
Finally, when $v_{\text{rep}}$ exceeds $v_{\text{att}}$, the steric repulsion is strong enough to overcome the underlying attractive forces at a pairwise level, resulting in uniform shielding of the underlying attractive interactions and recovery of the dispersed state.
The packing parameter-based argument has similarities with the geometric packing theory for aggregation of amphiphilic surfactants, where a dimensionless packing parameter relating the head and tail group shapes predicts formation of micelles and bilayers.\cite{Israelachvili2010}

\begin{figure*}
\centering
\includegraphics[width=0.90\textwidth]{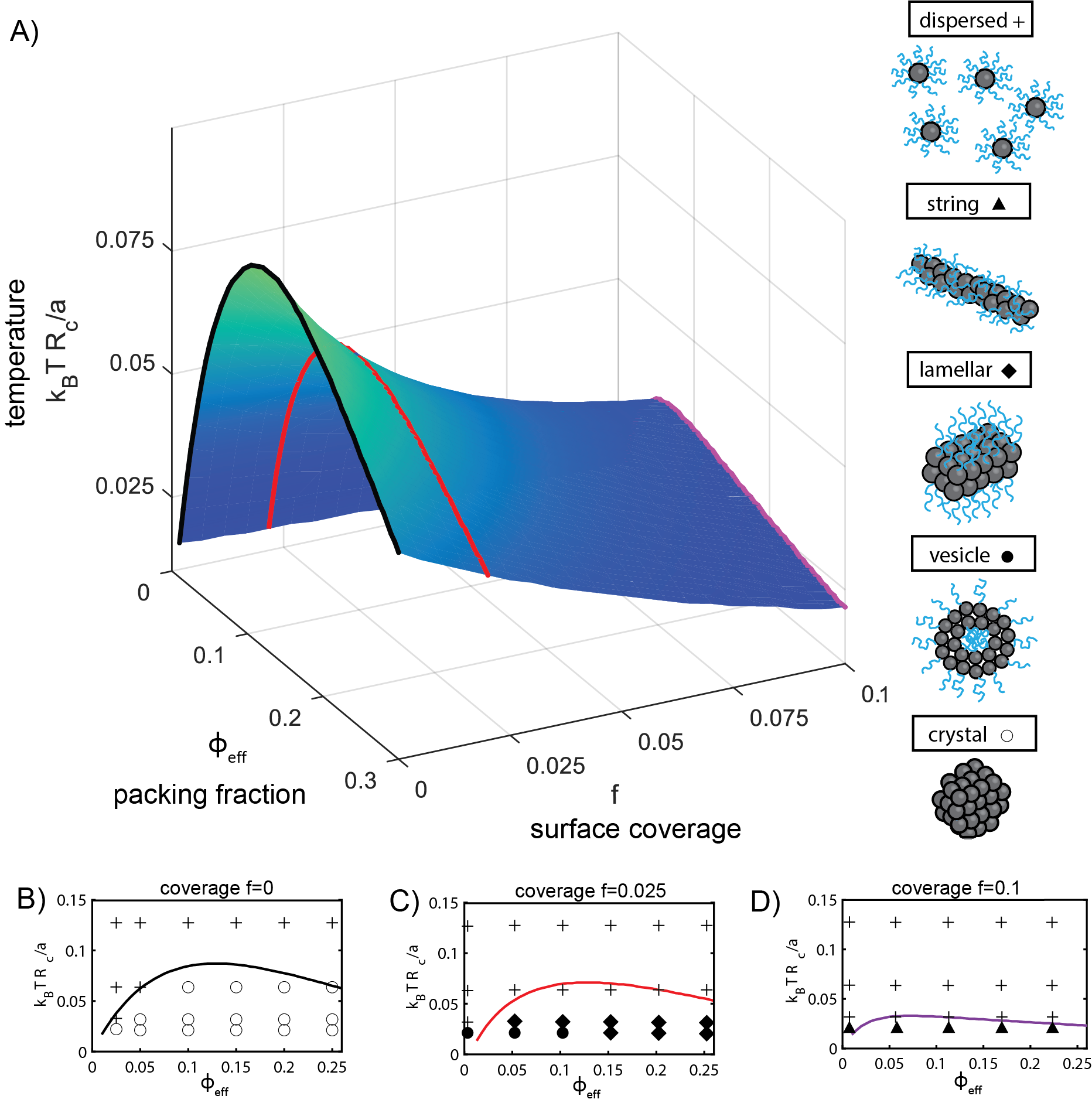}
\caption{Phase diagram of sticky colloids coated with surface-mobile polymers acting as dynamic surfactants. 
A) Three-dimensional spinodal curve demonstrating phase behavior with varying temperature, effective colloid packing fraction, and surface polymer coverage.
Color gradient varies with reduced temperature as a guide for the eye.
Cross-sectional plots for bare colloids (B) $f = 0$ (black) and fixed surface coverage of (C) $f = 0.025$ (red), (D) $f = 0.1$ (purple).
Solutions to our theory Eq.~\ref{OZ} are shown in solid lines and BD simulation state points are shown in markers.
}
\label{Fig3}
\end{figure*} 

\subsection{Self assembly phase diagram}
To gain further insight on how parameters such as temperature and polymer coverage regulate self assembly, we develop a simple theoretical model using a mean-field DDFT theory, which has been successfully used to model spinodal decomposition of simple colloidal fluids.\cite{Archer2004}
DDFT has also been applied to colloids interacting via long-ranged interactions to form repulsive crystals.\cite{VanTeeffelen2008}
Under the DDFT framework, one obtains the following linear stability result for the growth rate of instabilities perturbed about a uniform suspension of density, $\psi_{0}$ :
\begin{equation}\label{growth_rate}
    \omega = - D_{\text{c}} k^2 \left( 1 - \psi_{0} \frac{\hat{c}(k)}{k_{\text{B}}T} \right),
\end{equation}
where $\hat{c}(k)$ is the Fourier transform of the direct correlation function $c(r)$, which is related to the excess free energy in Eq.~\ref{Free_Energy}.\cite{Archer2004}
When the real part of the growth rate $\omega$ is negative for all wavenumbers $k$, the system is linearly stable against fluctuations.
In contrast, if the real part of $\omega$ is positive, phase separation via spinodal decomposition is predicted to occur at a wavenumber dominated by a critical value $k_{\text{max}}$.

In Fig.~\ref{Fig3}a, we present the spinodal curve as a function of the reduced temperature, effective packing fraction, and surface polymer coverage.
Because the presence of polymers slightly increases the 3D crowding, we report the effective packing fraction $\phi_{\text{eff}} = \phi + N_{\text{c}} m n \frac{\pi d_{\text{p}}^3}{6L^3}$ which corrects for the volume of the polymer beads.
Above the spinodal, the system is linearly stable against small fluctuations and exists as a homogenous suspension.
When quenched to a low reduced temperature below the spinodal, the system undergoes spontaneous phase separation.
We chose to vary the number of polymers on the colloidal surface $m$ to sweep different surface coverages $f$ in the simulations and keep the chain length $n$ constant.
Cross-sections at constant surface coverages $f$ are presented in Fig.~\ref{Fig3} b-d for better visualization, where we validate our theoretical results with BD data.
The dispersed phase corresponds to only a single peak in the radial distribution function $g(r)$, while all other phases are classified based on their morphologies which has been tabulated in Table~\ref{Table1}.

When no polymers are present ($f = 0$), we recover exactly the Archer and Evans result for the phase diagram, as shown in Fig.~\ref{Fig3}b.\cite{Archer2004}
The BD simulations exhibit either dispersed suspensions or tight, crystalline structures, and quantitatively agree with our theoretical result.
As $f$ increases, the polymers more effectively screen the attraction, and the spinodal becomes suppressed.
At surface coverages of $f=0.025$, we observe formation of vesicles in larger simulation boxes ($\phi_{\text{eff}} \leq 0.1$)  and lamellar phases in smaller boxes ($\phi_{\text{eff}} > 0.1)$.
At the highest surface coverages $f = 0.1$, we observed only the dispersed state and the string-like phase.
While morphologies observed previously in Table ~\ref{Table1} exactly correspond to regions in the spinodal, our theory is designed to predict the initial instability and onset of self assembly, not the late-time evolution into higher-order morphologies.
We attribute this limitation to the linear stability analysis, which only considers small disturbances to the homogeneous fluid. 
Additionally, by inputing only the dilute, two-body interactions $V(r)$ which are isotropic, we are unable to capture the anisotropic interactions that generate the higher order assemblies, as shown in Fig.~\ref{Fig2} and Table~\ref{Table1}.

In conclusion, we have demonstrated that sticky colloids coated with surface-mobile polymers exhibit rich self assembled morphologies due to coupling between macroscopic structure and microscopic polymer rearrangements.
We observed that surface-mobile polymers behave as ``dynamic surfactants'' that reorganize and promote anisotropic assembly of vesicle, lamellar, and string phases.
A critical geometric packing parameter describing the polymer free volume precisely governs structure formation, much like the theory for self assembly of amphiphilic surfactants.
Furthermore, our analytical theory describes the coupling between the reduced temperature and surface coverage required to observe self assembly via a phase-separation mechanism.
In contrast to traditional polymer-grafted particles whose polymer grafting sites are chemically fixed, surface mobility enables isotropic colloids to acquire different shapes during the assembly progress, which is particularly important in nonequilibrium settings.
Further work in this area could leverage nonequilibrium effects such as fluid flow to control the rate at which self assembly occurs, which may enable dynamic structural reconfiguration.
Hydrodynamic interactions also influence colloidal pair distributions as well as phase separation of colloidal gels.\cite{Tateno2021}
Surface mobile groups on emulsions and droplets have been shown to display interesting folding mechanisms during self assembly, which could be leveraged as well. \cite{McMullen2022,Mitra2023}

%% file: main.bbl
\begin{thebibliography}{45}%
\makeatletter
\providecommand \@ifxundefined [1]{%
 \@ifx{#1\undefined}
}%
\providecommand \@ifnum [1]{%
 \ifnum #1\expandafter \@firstoftwo
 \else \expandafter \@secondoftwo
 \fi
}%
\providecommand \@ifx [1]{%
 \ifx #1\expandafter \@firstoftwo
 \else \expandafter \@secondoftwo
 \fi
}%
\providecommand \natexlab [1]{#1}%
\providecommand \enquote  [1]{``#1''}%
\providecommand \bibnamefont  [1]{#1}%
\providecommand \bibfnamefont [1]{#1}%
\providecommand \citenamefont [1]{#1}%
\providecommand \href@noop [0]{\@secondoftwo}%
\providecommand \href [0]{\begingroup \@sanitize@url \@href}%
\providecommand \@href[1]{\@@startlink{#1}\@@href}%
\providecommand \@@href[1]{\endgroup#1\@@endlink}%
\providecommand \@sanitize@url [0]{\catcode `\\12\catcode `\$12\catcode `\&12\catcode `\#12\catcode `\^12\catcode `\_12\catcode `\%12\relax}%
\providecommand \@@startlink[1]{}%
\providecommand \@@endlink[0]{}%
\providecommand \url  [0]{\begingroup\@sanitize@url \@url }%
\providecommand \@url [1]{\endgroup\@href {#1}{\urlprefix }}%
\providecommand \urlprefix  [0]{URL }%
\providecommand \Eprint [0]{\href }%
\providecommand \doibase [0]{http://dx.doi.org/}%
\providecommand \selectlanguage [0]{\@gobble}%
\providecommand \bibinfo  [0]{\@secondoftwo}%
\providecommand \bibfield  [0]{\@secondoftwo}%
\providecommand \translation [1]{[#1]}%
\providecommand \BibitemOpen [0]{}%
\providecommand \bibitemStop [0]{}%
\providecommand \bibitemNoStop [0]{.\EOS\space}%
\providecommand \EOS [0]{\spacefactor3000\relax}%
\providecommand \BibitemShut  [1]{\csname bibitem#1\endcsname}%
\let\auto@bib@innerbib\@empty
\bibitem [{\citenamefont {Glotzer}\ \emph {et~al.}(2004)\citenamefont {Glotzer}, \citenamefont {Solomon},\ and\ \citenamefont {Kotov}}]{Glotzer2004}%
  \BibitemOpen
  \bibfield  {author} {\bibinfo {author} {\bibfnamefont {S.~C.}\ \bibnamefont {Glotzer}}, \bibinfo {author} {\bibfnamefont {M.~J.}\ \bibnamefont {Solomon}}, \ and\ \bibinfo {author} {\bibfnamefont {N.~A.}\ \bibnamefont {Kotov}},\ }\href {\doibase 10.1002/aic.10413} {\bibfield  {journal} {\bibinfo  {journal} {AIChE Journal}\ }\textbf {\bibinfo {volume} {50}},\ \bibinfo {pages} {2978} (\bibinfo {year} {2004})}\BibitemShut {NoStop}%
\bibitem [{\citenamefont {Sacanna}\ and\ \citenamefont {Pine}(2011)}]{Sacanna2011}%
  \BibitemOpen
  \bibfield  {author} {\bibinfo {author} {\bibfnamefont {S.}~\bibnamefont {Sacanna}}\ and\ \bibinfo {author} {\bibfnamefont {D.~J.}\ \bibnamefont {Pine}},\ }\href {\doibase 10.1016/j.cocis.2011.01.003} {\bibfield  {journal} {\bibinfo  {journal} {Current Opinion in Colloid and Interface Science}\ }\textbf {\bibinfo {volume} {16}},\ \bibinfo {pages} {96} (\bibinfo {year} {2011})}\BibitemShut {NoStop}%
\bibitem [{\citenamefont {Manoharan}(2015)}]{Manoharan2015}%
  \BibitemOpen
  \bibfield  {author} {\bibinfo {author} {\bibfnamefont {V.~N.}\ \bibnamefont {Manoharan}},\ }\href {\doibase 10.1126/science.1253751} {\bibfield  {journal} {\bibinfo  {journal} {Science}\ }\textbf {\bibinfo {volume} {349}} (\bibinfo {year} {2015}),\ 10.1126/science.1253751}\BibitemShut {NoStop}%
\bibitem [{\citenamefont {Sacanna}\ \emph {et~al.}(2013)\citenamefont {Sacanna}, \citenamefont {Korpics}, \citenamefont {Rodriguez}, \citenamefont {Col{\'{o}}n-Mel{\'{e}}ndez}, \citenamefont {Kim}, \citenamefont {Pine},\ and\ \citenamefont {Yi}}]{Sacanna2013}%
  \BibitemOpen
  \bibfield  {author} {\bibinfo {author} {\bibfnamefont {S.}~\bibnamefont {Sacanna}}, \bibinfo {author} {\bibfnamefont {M.}~\bibnamefont {Korpics}}, \bibinfo {author} {\bibfnamefont {K.}~\bibnamefont {Rodriguez}}, \bibinfo {author} {\bibfnamefont {L.}~\bibnamefont {Col{\'{o}}n-Mel{\'{e}}ndez}}, \bibinfo {author} {\bibfnamefont {S.-H.}\ \bibnamefont {Kim}}, \bibinfo {author} {\bibfnamefont {D.~J.}\ \bibnamefont {Pine}}, \ and\ \bibinfo {author} {\bibfnamefont {G.-R.}\ \bibnamefont {Yi}},\ }\href {\doibase 10.1038/ncomms2694} {\bibfield  {journal} {\bibinfo  {journal} {Nature Communications}\ }\textbf {\bibinfo {volume} {4}},\ \bibinfo {pages} {1688} (\bibinfo {year} {2013})}\BibitemShut {NoStop}%
\bibitem [{\citenamefont {van Anders}\ \emph {et~al.}(2014)\citenamefont {van Anders}, \citenamefont {Ahmed}, \citenamefont {Smith}, \citenamefont {Engel},\ and\ \citenamefont {Glotzer}}]{Anders2014}%
  \BibitemOpen
  \bibfield  {author} {\bibinfo {author} {\bibfnamefont {G.}~\bibnamefont {van Anders}}, \bibinfo {author} {\bibfnamefont {N.~K.}\ \bibnamefont {Ahmed}}, \bibinfo {author} {\bibfnamefont {R.}~\bibnamefont {Smith}}, \bibinfo {author} {\bibfnamefont {M.}~\bibnamefont {Engel}}, \ and\ \bibinfo {author} {\bibfnamefont {S.~C.}\ \bibnamefont {Glotzer}},\ }\href {\doibase 10.1021/nn4057353} {\bibfield  {journal} {\bibinfo  {journal} {ACS Nano}\ }\textbf {\bibinfo {volume} {8}},\ \bibinfo {pages} {931} (\bibinfo {year} {2014})}\BibitemShut {NoStop}%
\bibitem [{\citenamefont {Chen}\ \emph {et~al.}(2019)\citenamefont {Chen}, \citenamefont {Gibson}, \citenamefont {Liu}, \citenamefont {Rees}, \citenamefont {Lee}, \citenamefont {Xia}, \citenamefont {Lin}, \citenamefont {Xin}, \citenamefont {Gang},\ and\ \citenamefont {Weizmann}}]{Chen2019}%
  \BibitemOpen
  \bibfield  {author} {\bibinfo {author} {\bibfnamefont {G.}~\bibnamefont {Chen}}, \bibinfo {author} {\bibfnamefont {K.~J.}\ \bibnamefont {Gibson}}, \bibinfo {author} {\bibfnamefont {D.}~\bibnamefont {Liu}}, \bibinfo {author} {\bibfnamefont {H.~C.}\ \bibnamefont {Rees}}, \bibinfo {author} {\bibfnamefont {J.-H.}\ \bibnamefont {Lee}}, \bibinfo {author} {\bibfnamefont {W.}~\bibnamefont {Xia}}, \bibinfo {author} {\bibfnamefont {R.}~\bibnamefont {Lin}}, \bibinfo {author} {\bibfnamefont {H.~L.}\ \bibnamefont {Xin}}, \bibinfo {author} {\bibfnamefont {O.}~\bibnamefont {Gang}}, \ and\ \bibinfo {author} {\bibfnamefont {Y.}~\bibnamefont {Weizmann}},\ }\href {\doibase 10.1038/s41563-018-0231-1} {\bibfield  {journal} {\bibinfo  {journal} {Nature Materials}\ }\textbf {\bibinfo {volume} {18}},\ \bibinfo {pages} {169} (\bibinfo {year} {2019})}\BibitemShut {NoStop}%
\bibitem [{\citenamefont {Wang}\ \emph {et~al.}(2012)\citenamefont {Wang}, \citenamefont {Wang}, \citenamefont {Breed}, \citenamefont {Manoharan}, \citenamefont {Feng}, \citenamefont {Hollingsworth}, \citenamefont {Weck},\ and\ \citenamefont {Pine}}]{Wang2012}%
  \BibitemOpen
  \bibfield  {author} {\bibinfo {author} {\bibfnamefont {Y.}~\bibnamefont {Wang}}, \bibinfo {author} {\bibfnamefont {Y.}~\bibnamefont {Wang}}, \bibinfo {author} {\bibfnamefont {D.~R.}\ \bibnamefont {Breed}}, \bibinfo {author} {\bibfnamefont {V.~N.}\ \bibnamefont {Manoharan}}, \bibinfo {author} {\bibfnamefont {L.}~\bibnamefont {Feng}}, \bibinfo {author} {\bibfnamefont {A.~D.}\ \bibnamefont {Hollingsworth}}, \bibinfo {author} {\bibfnamefont {M.}~\bibnamefont {Weck}}, \ and\ \bibinfo {author} {\bibfnamefont {D.~J.}\ \bibnamefont {Pine}},\ }\href {\doibase 10.1038/nature11564} {\bibfield  {journal} {\bibinfo  {journal} {Nature}\ }\textbf {\bibinfo {volume} {491}},\ \bibinfo {pages} {51} (\bibinfo {year} {2012})}\BibitemShut {NoStop}%
\bibitem [{\citenamefont {Zhang}\ \emph {et~al.}(2021)\citenamefont {Zhang}, \citenamefont {Lyu}, \citenamefont {Xu}, \citenamefont {Mu},\ and\ \citenamefont {Wang}}]{Zhang2021}%
  \BibitemOpen
  \bibfield  {author} {\bibinfo {author} {\bibfnamefont {T.}~\bibnamefont {Zhang}}, \bibinfo {author} {\bibfnamefont {D.}~\bibnamefont {Lyu}}, \bibinfo {author} {\bibfnamefont {W.}~\bibnamefont {Xu}}, \bibinfo {author} {\bibfnamefont {Y.}~\bibnamefont {Mu}}, \ and\ \bibinfo {author} {\bibfnamefont {Y.}~\bibnamefont {Wang}},\ }\href {\doibase 10.3389/FPHY.2021.672375} {\bibfield  {journal} {\bibinfo  {journal} {Frontiers in Physics}\ }\textbf {\bibinfo {volume} {0}},\ \bibinfo {pages} {330} (\bibinfo {year} {2021})}\BibitemShut {NoStop}%
\bibitem [{\citenamefont {Angioletti-Uberti}\ \emph {et~al.}(2014)\citenamefont {Angioletti-Uberti}, \citenamefont {Varilly}, \citenamefont {Mognetti},\ and\ \citenamefont {Frenkel}}]{Angioletti-Uberti2014}%
  \BibitemOpen
  \bibfield  {author} {\bibinfo {author} {\bibfnamefont {S.}~\bibnamefont {Angioletti-Uberti}}, \bibinfo {author} {\bibfnamefont {P.}~\bibnamefont {Varilly}}, \bibinfo {author} {\bibfnamefont {B.~M.}\ \bibnamefont {Mognetti}}, \ and\ \bibinfo {author} {\bibfnamefont {D.}~\bibnamefont {Frenkel}},\ }\href {\doibase 10.1103/PhysRevLett.113.128303} {\bibfield  {journal} {\bibinfo  {journal} {Physical Review Letters}\ }\textbf {\bibinfo {volume} {113}},\ \bibinfo {pages} {128303} (\bibinfo {year} {2014})},\ \Eprint {http://arxiv.org/abs/1406.2870} {1406.2870} \BibitemShut {NoStop}%
\bibitem [{\citenamefont {Romano}\ \emph {et~al.}(2011)\citenamefont {Romano}, \citenamefont {Sanz},\ and\ \citenamefont {Sciortino}}]{Romano2011}%
  \BibitemOpen
  \bibfield  {author} {\bibinfo {author} {\bibfnamefont {F.}~\bibnamefont {Romano}}, \bibinfo {author} {\bibfnamefont {E.}~\bibnamefont {Sanz}}, \ and\ \bibinfo {author} {\bibfnamefont {F.}~\bibnamefont {Sciortino}},\ }\href {\doibase 10.1063/1.3578182} {\bibfield  {journal} {\bibinfo  {journal} {The Journal of Chemical Physics}\ }\textbf {\bibinfo {volume} {134}} (\bibinfo {year} {2011}),\ 10.1063/1.3578182}\BibitemShut {NoStop}%
\bibitem [{\citenamefont {Onsager}(1949)}]{Onsager1949}%
  \BibitemOpen
  \bibfield  {author} {\bibinfo {author} {\bibfnamefont {L.}~\bibnamefont {Onsager}},\ }\href {\doibase 10.1111/j.1749-6632.1949.tb27296.x} {\bibfield  {journal} {\bibinfo  {journal} {Annals of the New York Academy of Sciences}\ }\textbf {\bibinfo {volume} {51}},\ \bibinfo {pages} {627} (\bibinfo {year} {1949})}\BibitemShut {NoStop}%
\bibitem [{\citenamefont {Barry}\ and\ \citenamefont {Dogic}(2010)}]{Barry2010}%
  \BibitemOpen
  \bibfield  {author} {\bibinfo {author} {\bibfnamefont {E.}~\bibnamefont {Barry}}\ and\ \bibinfo {author} {\bibfnamefont {Z.}~\bibnamefont {Dogic}},\ }\href {\doibase 10.1073/pnas.1000406107} {\bibfield  {journal} {\bibinfo  {journal} {Proceedings of the National Academy of Sciences}\ }\textbf {\bibinfo {volume} {107}},\ \bibinfo {pages} {10348} (\bibinfo {year} {2010})}\BibitemShut {NoStop}%
\bibitem [{\citenamefont {Cinacchi}\ and\ \citenamefont {van Duijneveldt}(2010)}]{Cinacchi2010}%
  \BibitemOpen
  \bibfield  {author} {\bibinfo {author} {\bibfnamefont {G.}~\bibnamefont {Cinacchi}}\ and\ \bibinfo {author} {\bibfnamefont {J.~S.}\ \bibnamefont {van Duijneveldt}},\ }\href {\doibase 10.1021/jz900448e} {\bibfield  {journal} {\bibinfo  {journal} {The Journal of Physical Chemistry Letters}\ }\textbf {\bibinfo {volume} {1}},\ \bibinfo {pages} {787} (\bibinfo {year} {2010})}\BibitemShut {NoStop}%
\bibitem [{\citenamefont {Akcora}\ \emph {et~al.}(2009)\citenamefont {Akcora}, \citenamefont {Liu}, \citenamefont {Kumar}, \citenamefont {Moll}, \citenamefont {Li}, \citenamefont {Benicewicz}, \citenamefont {Schadler}, \citenamefont {Acehan}, \citenamefont {Panagiotopoulos}, \citenamefont {Pryamitsyn}, \citenamefont {Ganesan}, \citenamefont {Ilavsky}, \citenamefont {Thiyagarajan}, \citenamefont {Colby},\ and\ \citenamefont {Douglas}}]{Akcora2009}%
  \BibitemOpen
  \bibfield  {author} {\bibinfo {author} {\bibfnamefont {P.}~\bibnamefont {Akcora}}, \bibinfo {author} {\bibfnamefont {H.}~\bibnamefont {Liu}}, \bibinfo {author} {\bibfnamefont {S.~K.}\ \bibnamefont {Kumar}}, \bibinfo {author} {\bibfnamefont {J.}~\bibnamefont {Moll}}, \bibinfo {author} {\bibfnamefont {Y.}~\bibnamefont {Li}}, \bibinfo {author} {\bibfnamefont {B.~C.}\ \bibnamefont {Benicewicz}}, \bibinfo {author} {\bibfnamefont {L.~S.}\ \bibnamefont {Schadler}}, \bibinfo {author} {\bibfnamefont {D.}~\bibnamefont {Acehan}}, \bibinfo {author} {\bibfnamefont {A.~Z.}\ \bibnamefont {Panagiotopoulos}}, \bibinfo {author} {\bibfnamefont {V.}~\bibnamefont {Pryamitsyn}}, \bibinfo {author} {\bibfnamefont {V.}~\bibnamefont {Ganesan}}, \bibinfo {author} {\bibfnamefont {J.}~\bibnamefont {Ilavsky}}, \bibinfo {author} {\bibfnamefont {P.}~\bibnamefont {Thiyagarajan}}, \bibinfo {author} {\bibfnamefont {R.~H.}\ \bibnamefont {Colby}}, \ and\ \bibinfo {author} {\bibfnamefont {J.~F.}\ \bibnamefont {Douglas}},\ }\href {\doibase
  10.1038/nmat2404} {\bibfield  {journal} {\bibinfo  {journal} {Nature Materials}\ }\textbf {\bibinfo {volume} {8}},\ \bibinfo {pages} {354} (\bibinfo {year} {2009})}\BibitemShut {NoStop}%
\bibitem [{\citenamefont {Xu}\ \emph {et~al.}(2023)\citenamefont {Xu}, \citenamefont {Choi}, \citenamefont {Nagella},\ and\ \citenamefont {Takatori}}]{Xu2023}%
  \BibitemOpen
  \bibfield  {author} {\bibinfo {author} {\bibfnamefont {Y.}~\bibnamefont {Xu}}, \bibinfo {author} {\bibfnamefont {K.~H.}\ \bibnamefont {Choi}}, \bibinfo {author} {\bibfnamefont {S.~G.}\ \bibnamefont {Nagella}}, \ and\ \bibinfo {author} {\bibfnamefont {S.~C.}\ \bibnamefont {Takatori}},\ }\href {\doibase 10.1039/D3SM00673E} {\bibfield  {journal} {\bibinfo  {journal} {Soft Matter}\ }\textbf {\bibinfo {volume} {19}},\ \bibinfo {pages} {5692} (\bibinfo {year} {2023})}\BibitemShut {NoStop}%
\bibitem [{\citenamefont {Leunissen}\ \emph {et~al.}(2009)\citenamefont {Leunissen}, \citenamefont {Dreyfus}, \citenamefont {Cheong}, \citenamefont {Grier}, \citenamefont {Sha}, \citenamefont {Seeman},\ and\ \citenamefont {Chaikin}}]{Leunissen2009}%
  \BibitemOpen
  \bibfield  {author} {\bibinfo {author} {\bibfnamefont {M.~E.}\ \bibnamefont {Leunissen}}, \bibinfo {author} {\bibfnamefont {R.}~\bibnamefont {Dreyfus}}, \bibinfo {author} {\bibfnamefont {F.~C.}\ \bibnamefont {Cheong}}, \bibinfo {author} {\bibfnamefont {D.~G.}\ \bibnamefont {Grier}}, \bibinfo {author} {\bibfnamefont {R.}~\bibnamefont {Sha}}, \bibinfo {author} {\bibfnamefont {N.~C.}\ \bibnamefont {Seeman}}, \ and\ \bibinfo {author} {\bibfnamefont {P.~M.}\ \bibnamefont {Chaikin}},\ }\href {\doibase 10.1038/nmat2471} {\bibfield  {journal} {\bibinfo  {journal} {Nature Materials}\ }\textbf {\bibinfo {volume} {8}},\ \bibinfo {pages} {590} (\bibinfo {year} {2009})}\BibitemShut {NoStop}%
\bibitem [{\citenamefont {{Van Der Meulen}}\ and\ \citenamefont {Leunissen}(2013)}]{VanDerMeulen2013}%
  \BibitemOpen
  \bibfield  {author} {\bibinfo {author} {\bibfnamefont {S.~A.}\ \bibnamefont {{Van Der Meulen}}}\ and\ \bibinfo {author} {\bibfnamefont {M.~E.}\ \bibnamefont {Leunissen}},\ }\href {\doibase 10.1021/ja406226b} {\bibfield  {journal} {\bibinfo  {journal} {Journal of the American Chemical Society}\ }\textbf {\bibinfo {volume} {135}},\ \bibinfo {pages} {15129} (\bibinfo {year} {2013})}\BibitemShut {NoStop}%
\bibitem [{\citenamefont {Mitra}\ \emph {et~al.}(2023)\citenamefont {Mitra}, \citenamefont {Chang}, \citenamefont {McMullen}, \citenamefont {Puchall}, \citenamefont {Brujic},\ and\ \citenamefont {Hocky}}]{Mitra2023}%
  \BibitemOpen
  \bibfield  {author} {\bibinfo {author} {\bibfnamefont {G.}~\bibnamefont {Mitra}}, \bibinfo {author} {\bibfnamefont {C.}~\bibnamefont {Chang}}, \bibinfo {author} {\bibfnamefont {A.}~\bibnamefont {McMullen}}, \bibinfo {author} {\bibfnamefont {D.}~\bibnamefont {Puchall}}, \bibinfo {author} {\bibfnamefont {J.}~\bibnamefont {Brujic}}, \ and\ \bibinfo {author} {\bibfnamefont {G.~M.}\ \bibnamefont {Hocky}},\ }\href {\doibase 10.1039/D3SM00196B} {\bibfield  {journal} {\bibinfo  {journal} {Soft Matter}\ }\textbf {\bibinfo {volume} {19}},\ \bibinfo {pages} {4223} (\bibinfo {year} {2023})}\BibitemShut {NoStop}%
\bibitem [{\citenamefont {Israelachvili}(2010)}]{Israelachvili2010}%
  \BibitemOpen
  \bibfield  {author} {\bibinfo {author} {\bibfnamefont {J.~N.}\ \bibnamefont {Israelachvili}},\ }\href {\doibase 10.1016/C2009-0-21560-1} {\bibfield  {journal} {\bibinfo  {journal} {Intermolecular and Surface Forces, Third Edition}\ ,\ \bibinfo {pages} {1}} (\bibinfo {year} {2010})}\BibitemShut {NoStop}%
\bibitem [{\citenamefont {Kremer}\ and\ \citenamefont {Grest}(1990)}]{Kremer1990}%
  \BibitemOpen
  \bibfield  {author} {\bibinfo {author} {\bibfnamefont {K.}~\bibnamefont {Kremer}}\ and\ \bibinfo {author} {\bibfnamefont {G.~S.}\ \bibnamefont {Grest}},\ }\href {\doibase 10.1063/1.458541} {\bibfield  {journal} {\bibinfo  {journal} {The Journal of Chemical Physics}\ }\textbf {\bibinfo {volume} {92}},\ \bibinfo {pages} {5057} (\bibinfo {year} {1990})}\BibitemShut {NoStop}%
\bibitem [{\citenamefont {Anderson}\ \emph {et~al.}(2020)\citenamefont {Anderson}, \citenamefont {Glaser},\ and\ \citenamefont {Glotzer}}]{Anderson2020}%
  \BibitemOpen
  \bibfield  {author} {\bibinfo {author} {\bibfnamefont {J.~A.}\ \bibnamefont {Anderson}}, \bibinfo {author} {\bibfnamefont {J.}~\bibnamefont {Glaser}}, \ and\ \bibinfo {author} {\bibfnamefont {S.~C.}\ \bibnamefont {Glotzer}},\ }\href {\doibase 10.1016/j.commatsci.2019.109363} {\bibfield  {journal} {\bibinfo  {journal} {Computational Materials Science}\ }\textbf {\bibinfo {volume} {173}},\ \bibinfo {pages} {109363} (\bibinfo {year} {2020})}\BibitemShut {NoStop}%
\bibitem [{\citenamefont {Hansen}\ and\ \citenamefont {McDonald}(2013)}]{Hansen2013}%
  \BibitemOpen
  \bibfield  {author} {\bibinfo {author} {\bibfnamefont {J.~P.}\ \bibnamefont {Hansen}}\ and\ \bibinfo {author} {\bibfnamefont {I.~R.}\ \bibnamefont {McDonald}},\ }\href {\doibase 10.1016/C2010-0-66723-X} {\emph {\bibinfo {title} {Theory of Simple Liquids: With Applications to Soft Matter: Fourth Edition}}}\ (\bibinfo  {publisher} {Elsevier},\ \bibinfo {year} {2013})\ pp.\ \bibinfo {pages} {1--619}\BibitemShut {NoStop}%
\bibitem [{\citenamefont {de~Gennes}(1987)}]{DeGennes1987}%
  \BibitemOpen
  \bibfield  {author} {\bibinfo {author} {\bibfnamefont {P.~G.}\ \bibnamefont {de~Gennes}},\ }\href {\doibase 10.1016/0001-8686(87)85003-0} {\enquote {\bibinfo {title} {{Polymers at an interface; a simplified view}},}\ } (\bibinfo {year} {1987})\BibitemShut {NoStop}%
\bibitem [{\citenamefont {Milner}(1991)}]{Milner1991}%
  \BibitemOpen
  \bibfield  {author} {\bibinfo {author} {\bibfnamefont {S.~T.}\ \bibnamefont {Milner}},\ }\href {\doibase 10.1126/science.251.4996.905} {\bibfield  {journal} {\bibinfo  {journal} {Science}\ }\textbf {\bibinfo {volume} {251}},\ \bibinfo {pages} {905} (\bibinfo {year} {1991})}\BibitemShut {NoStop}%
\bibitem [{\citenamefont {Hecht}\ and\ \citenamefont {Hecht}(2012)}]{Hecht2012}%
  \BibitemOpen
  \bibfield  {author} {\bibinfo {author} {\bibfnamefont {F.}~\bibnamefont {Hecht}}\ and\ \bibinfo {author} {\bibfnamefont {F.}~\bibnamefont {Hecht}},\ }\href {\doibase 10.1515/jnum-2012-0013ï} {\bibfield  {journal} {\bibinfo  {journal} {Journal of Numerical Mathematics}\ }\textbf {\bibinfo {volume} {20}},\ \bibinfo {pages} {1} (\bibinfo {year} {2012})}\BibitemShut {NoStop}%
\bibitem [{\citenamefont {Loverso}\ \emph {et~al.}(2012)\citenamefont {Loverso}, \citenamefont {Egorov},\ and\ \citenamefont {Binder}}]{Loverso2012}%
  \BibitemOpen
  \bibfield  {author} {\bibinfo {author} {\bibfnamefont {F.}~\bibnamefont {Loverso}}, \bibinfo {author} {\bibfnamefont {S.~A.}\ \bibnamefont {Egorov}}, \ and\ \bibinfo {author} {\bibfnamefont {K.}~\bibnamefont {Binder}},\ }\href {\doibase 10.1021/ma301651z} {\bibfield  {journal} {\bibinfo  {journal} {Macromolecules}\ }\textbf {\bibinfo {volume} {45}},\ \bibinfo {pages} {8892} (\bibinfo {year} {2012})}\BibitemShut {NoStop}%
\bibitem [{\citenamefont {Caccamo}\ and\ \citenamefont {Giunta}(1993)}]{Caccamo1993}%
  \BibitemOpen
  \bibfield  {author} {\bibinfo {author} {\bibfnamefont {C.}~\bibnamefont {Caccamo}}\ and\ \bibinfo {author} {\bibfnamefont {G.}~\bibnamefont {Giunta}},\ }\href {\doibase 10.1080/00268979300100081} {\bibfield  {journal} {\bibinfo  {journal} {Molecular Physics}\ }\textbf {\bibinfo {volume} {78}},\ \bibinfo {pages} {83} (\bibinfo {year} {1993})}\BibitemShut {NoStop}%
\bibitem [{\citenamefont {Mahynski}\ and\ \citenamefont {Panagiotopoulos}(2015)}]{Mahynski2015}%
  \BibitemOpen
  \bibfield  {author} {\bibinfo {author} {\bibfnamefont {N.~A.}\ \bibnamefont {Mahynski}}\ and\ \bibinfo {author} {\bibfnamefont {A.~Z.}\ \bibnamefont {Panagiotopoulos}},\ }\href {\doibase 10.1063/1.4908044} {\bibfield  {journal} {\bibinfo  {journal} {The Journal of Chemical Physics}\ }\textbf {\bibinfo {volume} {142}} (\bibinfo {year} {2015}),\ 10.1063/1.4908044}\BibitemShut {NoStop}%
\bibitem [{\citenamefont {Rogers}(1980)}]{Rogers1980}%
  \BibitemOpen
  \bibfield  {author} {\bibinfo {author} {\bibfnamefont {F.~J.}\ \bibnamefont {Rogers}},\ }\href {\doibase 10.1063/1.440124} {\bibfield  {journal} {\bibinfo  {journal} {The Journal of Chemical Physics}\ }\textbf {\bibinfo {volume} {73}},\ \bibinfo {pages} {6272} (\bibinfo {year} {1980})}\BibitemShut {NoStop}%
\bibitem [{\citenamefont {Rogers}\ and\ \citenamefont {Young}(1984)}]{Rogers1984}%
  \BibitemOpen
  \bibfield  {author} {\bibinfo {author} {\bibfnamefont {F.~J.}\ \bibnamefont {Rogers}}\ and\ \bibinfo {author} {\bibfnamefont {D.~A.}\ \bibnamefont {Young}},\ }\href {\doibase 10.1103/PhysRevA.30.999} {\bibfield  {journal} {\bibinfo  {journal} {Physical Review A}\ }\textbf {\bibinfo {volume} {30}},\ \bibinfo {pages} {999} (\bibinfo {year} {1984})}\BibitemShut {NoStop}%
\bibitem [{\citenamefont {Archer}\ and\ \citenamefont {Evans}(2004)}]{Archer2004}%
  \BibitemOpen
  \bibfield  {author} {\bibinfo {author} {\bibfnamefont {A.~J.}\ \bibnamefont {Archer}}\ and\ \bibinfo {author} {\bibfnamefont {R.}~\bibnamefont {Evans}},\ }\href {\doibase 10.1063/1.1778374} {\bibfield  {journal} {\bibinfo  {journal} {The Journal of Chemical Physics}\ }\textbf {\bibinfo {volume} {121}},\ \bibinfo {pages} {4246} (\bibinfo {year} {2004})},\ \Eprint {http://arxiv.org/abs/0405665} {0405665 [cond-mat]} \BibitemShut {NoStop}%
\bibitem [{\citenamefont {Marconi}\ and\ \citenamefont {Tarazona}(1999)}]{Marconi1999}%
  \BibitemOpen
  \bibfield  {author} {\bibinfo {author} {\bibfnamefont {U.~M.~B.}\ \bibnamefont {Marconi}}\ and\ \bibinfo {author} {\bibfnamefont {P.}~\bibnamefont {Tarazona}},\ }\href {\doibase 10.1063/1.478705} {\bibfield  {journal} {\bibinfo  {journal} {The Journal of Chemical Physics}\ }\textbf {\bibinfo {volume} {110}},\ \bibinfo {pages} {8032} (\bibinfo {year} {1999})},\ \Eprint {http://arxiv.org/abs/9810403} {9810403 [cond-mat]} \BibitemShut {NoStop}%
\bibitem [{\citenamefont {Doi}\ \emph {et~al.}(1988)\citenamefont {Doi}, \citenamefont {Edwards},\ and\ \citenamefont {Edwards}}]{Doi1988}%
  \BibitemOpen
  \bibfield  {author} {\bibinfo {author} {\bibfnamefont {M.}~\bibnamefont {Doi}}, \bibinfo {author} {\bibfnamefont {S.~F.}\ \bibnamefont {Edwards}}, \ and\ \bibinfo {author} {\bibfnamefont {S.~F.}\ \bibnamefont {Edwards}},\ }\href@noop {} {\emph {\bibinfo {title} {The theory of polymer dynamics}}},\ Vol.~\bibinfo {volume} {73}\ (\bibinfo  {publisher} {oxford university press},\ \bibinfo {year} {1988})\BibitemShut {NoStop}%
\bibitem [{\citenamefont {Miller}\ and\ \citenamefont {Cacciuto}(2009)}]{Miller2009}%
  \BibitemOpen
  \bibfield  {author} {\bibinfo {author} {\bibfnamefont {W.~L.}\ \bibnamefont {Miller}}\ and\ \bibinfo {author} {\bibfnamefont {A.}~\bibnamefont {Cacciuto}},\ }\href {\doibase 10.1103/PhysRevE.80.021404} {\bibfield  {journal} {\bibinfo  {journal} {Physical Review E}\ }\textbf {\bibinfo {volume} {80}},\ \bibinfo {pages} {021404} (\bibinfo {year} {2009})}\BibitemShut {NoStop}%
\bibitem [{\citenamefont {Preisler}\ \emph {et~al.}(2014)\citenamefont {Preisler}, \citenamefont {Vissers}, \citenamefont {Muna{\`{o}}}, \citenamefont {Smallenburg},\ and\ \citenamefont {Sciortino}}]{Preisler2014}%
  \BibitemOpen
  \bibfield  {author} {\bibinfo {author} {\bibfnamefont {Z.}~\bibnamefont {Preisler}}, \bibinfo {author} {\bibfnamefont {T.}~\bibnamefont {Vissers}}, \bibinfo {author} {\bibfnamefont {G.}~\bibnamefont {Muna{\`{o}}}}, \bibinfo {author} {\bibfnamefont {F.}~\bibnamefont {Smallenburg}}, \ and\ \bibinfo {author} {\bibfnamefont {F.}~\bibnamefont {Sciortino}},\ }\href {\doibase 10.1039/c4sm00505h} {\bibfield  {journal} {\bibinfo  {journal} {Soft Matter}\ }\textbf {\bibinfo {volume} {10}},\ \bibinfo {pages} {5121} (\bibinfo {year} {2014})}\BibitemShut {NoStop}%
\bibitem [{\citenamefont {Hristova}\ and\ \citenamefont {Needham}(1994)}]{Hristova1994}%
  \BibitemOpen
  \bibfield  {author} {\bibinfo {author} {\bibfnamefont {K.}~\bibnamefont {Hristova}}\ and\ \bibinfo {author} {\bibfnamefont {D.}~\bibnamefont {Needham}},\ }\href {\doibase 10.1006/jcis.1994.1424} {\bibfield  {journal} {\bibinfo  {journal} {Journal of Colloid and Interface Science}\ }\textbf {\bibinfo {volume} {168}},\ \bibinfo {pages} {302} (\bibinfo {year} {1994})}\BibitemShut {NoStop}%
\bibitem [{\citenamefont {Segr{\`{e}}}\ \emph {et~al.}(2001)\citenamefont {Segr{\`{e}}}, \citenamefont {Prasad}, \citenamefont {Schofield},\ and\ \citenamefont {Weitz}}]{Segre2001}%
  \BibitemOpen
  \bibfield  {author} {\bibinfo {author} {\bibfnamefont {P.}~\bibnamefont {Segr{\`{e}}}}, \bibinfo {author} {\bibfnamefont {V.}~\bibnamefont {Prasad}}, \bibinfo {author} {\bibfnamefont {A.}~\bibnamefont {Schofield}}, \ and\ \bibinfo {author} {\bibfnamefont {D.}~\bibnamefont {Weitz}},\ }\href {\doibase 10.1103/PhysRevLett.86.6042} {\bibfield  {journal} {\bibinfo  {journal} {Physical Review Letters}\ }\textbf {\bibinfo {volume} {86}},\ \bibinfo {pages} {6042} (\bibinfo {year} {2001})}\BibitemShut {NoStop}%
\bibitem [{\citenamefont {Zia}\ \emph {et~al.}(2014)\citenamefont {Zia}, \citenamefont {Landrum},\ and\ \citenamefont {Russel}}]{Zia2014}%
  \BibitemOpen
  \bibfield  {author} {\bibinfo {author} {\bibfnamefont {R.~N.}\ \bibnamefont {Zia}}, \bibinfo {author} {\bibfnamefont {B.~J.}\ \bibnamefont {Landrum}}, \ and\ \bibinfo {author} {\bibfnamefont {W.~B.}\ \bibnamefont {Russel}},\ }\href {\doibase 10.1122/1.4892115} {\bibfield  {journal} {\bibinfo  {journal} {Citation: Journal of Rheology}\ }\textbf {\bibinfo {volume} {58}},\ \bibinfo {pages} {1121} (\bibinfo {year} {2014})}\BibitemShut {NoStop}%
\bibitem [{\citenamefont {Cooke}\ and\ \citenamefont {Deserno}(2005)}]{Cooke2005}%
  \BibitemOpen
  \bibfield  {author} {\bibinfo {author} {\bibfnamefont {I.~R.}\ \bibnamefont {Cooke}}\ and\ \bibinfo {author} {\bibfnamefont {M.}~\bibnamefont {Deserno}},\ }\href {\doibase 10.1063/1.2135785} {\bibfield  {journal} {\bibinfo  {journal} {The Journal of Chemical Physics}\ }\textbf {\bibinfo {volume} {123}} (\bibinfo {year} {2005}),\ 10.1063/1.2135785}\BibitemShut {NoStop}%
\bibitem [{\citenamefont {Yuan}\ \emph {et~al.}(2010)\citenamefont {Yuan}, \citenamefont {Huang}, \citenamefont {Li}, \citenamefont {Lykotrafitis},\ and\ \citenamefont {Zhang}}]{Yuan2010}%
  \BibitemOpen
  \bibfield  {author} {\bibinfo {author} {\bibfnamefont {H.}~\bibnamefont {Yuan}}, \bibinfo {author} {\bibfnamefont {C.}~\bibnamefont {Huang}}, \bibinfo {author} {\bibfnamefont {J.}~\bibnamefont {Li}}, \bibinfo {author} {\bibfnamefont {G.}~\bibnamefont {Lykotrafitis}}, \ and\ \bibinfo {author} {\bibfnamefont {S.}~\bibnamefont {Zhang}},\ }\href {\doibase 10.1103/PhysRevE.82.011905} {\bibfield  {journal} {\bibinfo  {journal} {Physical Review E - Statistical, Nonlinear, and Soft Matter Physics}\ }\textbf {\bibinfo {volume} {82}} (\bibinfo {year} {2010}),\ 10.1103/PhysRevE.82.011905}\BibitemShut {NoStop}%
\bibitem [{\citenamefont {Noguchi}(2011)}]{Noguchi2011}%
  \BibitemOpen
  \bibfield  {author} {\bibinfo {author} {\bibfnamefont {H.}~\bibnamefont {Noguchi}},\ }\href {\doibase 10.1063/1.3541246} {\bibfield  {journal} {\bibinfo  {journal} {The Journal of Chemical Physics}\ }\textbf {\bibinfo {volume} {134}} (\bibinfo {year} {2011}),\ 10.1063/1.3541246}\BibitemShut {NoStop}%
\bibitem [{\citenamefont {Shiba}\ and\ \citenamefont {Noguchi}(2011)}]{Shiba2011}%
  \BibitemOpen
  \bibfield  {author} {\bibinfo {author} {\bibfnamefont {H.}~\bibnamefont {Shiba}}\ and\ \bibinfo {author} {\bibfnamefont {H.}~\bibnamefont {Noguchi}},\ }\href {\doibase 10.1103/PhysRevE.84.031926} {\bibfield  {journal} {\bibinfo  {journal} {Physical Review E}\ }\textbf {\bibinfo {volume} {84}},\ \bibinfo {pages} {031926} (\bibinfo {year} {2011})}\BibitemShut {NoStop}%
\bibitem [{\citenamefont {{Van Teeffelen}}\ \emph {et~al.}(2008)\citenamefont {{Van Teeffelen}}, \citenamefont {Likos},\ and\ \citenamefont {L{\"{o}}wen}}]{VanTeeffelen2008}%
  \BibitemOpen
  \bibfield  {author} {\bibinfo {author} {\bibfnamefont {S.}~\bibnamefont {{Van Teeffelen}}}, \bibinfo {author} {\bibfnamefont {C.~N.}\ \bibnamefont {Likos}}, \ and\ \bibinfo {author} {\bibfnamefont {H.}~\bibnamefont {L{\"{o}}wen}},\ }\href {\doibase 10.1103/PHYSREVLETT.100.108302/FIGURES/5/MEDIUM} {\bibfield  {journal} {\bibinfo  {journal} {Physical Review Letters}\ }\textbf {\bibinfo {volume} {100}},\ \bibinfo {pages} {108302} (\bibinfo {year} {2008})},\ \Eprint {http://arxiv.org/abs/0802.2235} {arXiv:0802.2235} \BibitemShut {NoStop}%
\bibitem [{\citenamefont {Tateno}\ and\ \citenamefont {Tanaka}(2021)}]{Tateno2021}%
  \BibitemOpen
  \bibfield  {author} {\bibinfo {author} {\bibfnamefont {M.}~\bibnamefont {Tateno}}\ and\ \bibinfo {author} {\bibfnamefont {H.}~\bibnamefont {Tanaka}},\ }\href {\doibase 10.1038/s41467-020-20734-8} {\bibfield  {journal} {\bibinfo  {journal} {Nature Communications}\ }\textbf {\bibinfo {volume} {12}},\ \bibinfo {pages} {912} (\bibinfo {year} {2021})}\BibitemShut {NoStop}%
\bibitem [{\citenamefont {McMullen}\ \emph {et~al.}(2022)\citenamefont {McMullen}, \citenamefont {{Mu{\~{n}}oz Basagoiti}}, \citenamefont {Zeravcic},\ and\ \citenamefont {Brujic}}]{McMullen2022}%
  \BibitemOpen
  \bibfield  {author} {\bibinfo {author} {\bibfnamefont {A.}~\bibnamefont {McMullen}}, \bibinfo {author} {\bibfnamefont {M.}~\bibnamefont {{Mu{\~{n}}oz Basagoiti}}}, \bibinfo {author} {\bibfnamefont {Z.}~\bibnamefont {Zeravcic}}, \ and\ \bibinfo {author} {\bibfnamefont {J.}~\bibnamefont {Brujic}},\ }\href {\doibase 10.1038/s41586-022-05198-8} {\bibfield  {journal} {\bibinfo  {journal} {Nature}\ }\textbf {\bibinfo {volume} {610}},\ \bibinfo {pages} {502} (\bibinfo {year} {2022})}\BibitemShut {NoStop}%
\end{thebibliography}%
